# On Kalman-Like Finite Impulse Response Filters

Lubin Chang, *Member, IEEE*

Naval University of Engineering, Wuhan, China (e-mail:changlubin@163.com)

*Abstract*—This note reveals an explicit relationship between two representative finite impulse response (FIR) filters, i.e. the newly derived and popularized Kalman-Like unbiased FIR filter (UFIR) and the receding horizon Kalman FIR filter (RHKF). It is pointed out that the only difference of the two algorithms lies in the noise statistics ignorance and appropriate initial condition construction strategy in UFIR. The revelation can benefit the performance improvement of one by drawing lessons from the other. Some interesting conclusions have also been drawn and discussed from this revelation.

I. Introduction

Probabilistic inference or filtering is concerned with estimation of the dynamic state from noisy or incomplete observations. By far the primary mechanism historically used to make probabilistic inference has been the Kalman filter (KF). The KF is the optimal estimator when the noise is white Gaussian [1, 2]. The inherent virtue of KF has generated an enormous number of works devoted to its investigations and applications. However, the KF suffers from high sensitivity to modeling errors due to its infinite impulse response (IIR) nature. Here, the term *IIR* refers to the fact that all the measurements prior to the current time have effect on the state estimate at current time in the KF [3, 4]. In this case, the finite impulse response (FIR) estimator has been approved as a significant rival of the KF [5-17].

This work was supported in part by the National Natural Science Foundation of China (61304241, 61374206).

Naval University of Engineering, Wuhan 430033, China (e-mail:changlubin@163.com, Lubin.Chang.1987@ieee.org).

The FIR filters have a moving horizontal structure in which most recent past measurements are exploited. Despite the inherent virtue in terms of stability and robustness, FIR filters have drawn little attention in state estimation probably due to their analytical complexity and large computational burden [14].Toward improving the level of knowledge in this field, Kwon et al. derived a receding horizon Kalman FIR filter (RHKF) by combining the KF and the receding horizon strategy [7-9]. An iterative form of the RHKF filter has also been presented based on the one-step predicted information estimator for state estimation. The optimality, unbiasedness and deadbeat properties of RHKF have been demonstrated in [9]. Recently few years, Shmaliy et al. have infused new vitality into the studies of FIR for state estimation by investigating a series of iterative Kalman-Like unbiased FIR (UFIR) algorithms [10-17]. The term *Kalman-Like* means that these algorithms have the same predictor/corrector structure as the KF. The term *iterative* actually refers to the *recursive* utilization of all measurements on the finite interval. The procedure used to circumvent the KF's sensitive problem in UFIR is to totally ignores the statistics of the noise and the initial state.

Based on the aforementioned discussion, it can be found that both RHKF and UFIR are types of FIR with close connection with the KF. However, the explicit relationship between the two algorithms has yet not been revealed and discussed. This is mainly because that the iterative form of RHKF does not have an explicit predictor/corrector structure although it is derived through a modification of the KF. These facts represent the main motivation of this note, which will focus on presenting a predictor/corrector structure of the RHKF and revealing the explicit relationship between RHKF and UFIR. The predictor/corrector-structure perspective on the RHKF reveals that the UFIR can be readily obtained from RHKF by completely ignoring the noise statistics when the horizon initial state is not considered. This revelation provides mutual benefit for the two algorithms, that is, some fruitful strategies and methodologies in each scope can be applied to the improvement of the other.

II. Main Results

In this note, we shall consider the discrete time-variant linear model represented in state-space with

$$x_k = F_k x_{k-1} + w_{k-1} \quad (1a)$$

$$y_k = H_k x_k + v_k \quad (1b)$$

where $x_k \in \mathbb{R}^n$ is the state, $y_k \in \mathbb{R}^m$ is the measurement, $w_k \sim N(0, Q_k)$ is the process noise, $v_k \sim N(0, R_k)$ is the measurement noise. They are assumed to be independent of each other. The matrix $F_k \in \mathbb{R}^{n \times n}$ is the transition matrix of the dynamic model and $H_k \in \mathbb{R}^{m \times n}$ is the measurement model matrix.

In this section, the filtering equations for the linear filtering model (1) evaluated by KF, RHKF and UFIR with predictor/corrector structure are firstly presented, respectively. Then, their explicit relationships between each other are discussed and some interesting remarks are drawn.

*A. KF*

The KF is the closed form solution to the Bayesian filtering equations for the linear filtering model (1) with the following prediction and correction steps.

Given the initial estimate

$$\hat{x}_0 = E(x_0)$$

$$P_0 = E\left[(x_0 - \hat{x}_0)(x_0 - \hat{x}_0)^T\right]$$

For $k = 1, 2, \cdots$

The prediction step *is*

$$\hat{x}_{k|k-1} = F_k \hat{x}_{k-1} \quad (2)$$

$$P_{k|k-1} = F_k P_{k-1} F_k^T + Q_{k-1} \quad (3)$$

The correction step *is*

$$K_k = P_{k|k-1} H_k^T \left( H_k P_{k|k-1} H_k^T + R_k \right)^{-1} = P_k H_k^T R_k^{-1} \quad (4)$$

$$\hat{x}_k = \hat{x}_{k|k-1} + K_k \left( y_k - H_k \hat{x}_{k|k-1} \right) \quad (5)$$

$$P_k = P_{k|k-1} - K_k H_k P_{k|k-1} \quad (6)$$

The illustration of the KF recursion is shown in Fig. 1.

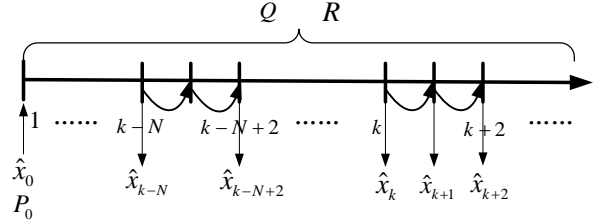

Fig. 1. Illustration of the KF Recursion

*B. RHKF in Predictor/corrector Structure*

The iterative (or recursive) RHKF is derived by embedding the KF into the moving horizontal structure. It can be found in [8] that in each moving horizontal interval of the RHKF, a one-step predicted information estimator is implemented recursively for the state estimation. In this respect, we can also formulate the iterative RHKF by the general information filter with the predictor/corrector structure. The resulting filtering algorithm can be described as follows.

Given a specified horizontal interval $N$ and the initial estimate of a certain moving horizontal interval $[k-N, k]$

$$Z_{k-N} = P_{k-N}^{-1} = 0_{n \times n}$$

$$\hat{z}_{k-N} = P_{k-N}^{-1} \hat{x}_{k-N} = 0_{n \times 1}$$

For $l = k - N + 1, k - N + 2, \cdots, k$

The prediction step *is*

$$\hat{z}_{l|l-1} = P_{l|l-1}^{-1} \hat{x}_{l|l-1} = P_{l|l-1}^{-1} F_l P_{l-1} \hat{z}_{l-1} = Z_{l|l-1} F_l Z_{l-1}^{-1} \hat{z}_{l-1} \quad (7)$$

$$Z_{l|l-1} = P_{l|l-1}^{-1} = \left( F_l P_{l-1} F_l^T + Q_{l-1} \right)^{-1} = \left( F_l Z_{l-1}^{-1} F_l^T + Q_{l-1} \right)^{-1} \quad (8)$$

The correction step *is*

$$K_l = P_l H_l^T R_l^{-1} \quad (9)$$

$$\hat{z}_l = P_l^{-1}\hat{x}_l = P_l^{-1}\left[\hat{x}_{l|l-1} + K_l\left(y_l - H_l\hat{x}_{l|l-1}\right)\right]$$
$$= \left(P_l^{-1} - H_l^T R_l^{-1} H_l\right)\hat{x}_{l|l-1} + H_l^T R_l^{-1} y_l \quad (10)$$
$$= \hat{z}_{l|l-1} + H_l^T R_l^{-1} y_l$$

$$Z_l = P_l^{-1} = P_{l|l-1}^{-1} + H_l^T R_l^{-1} H_l = Z_{l|l-1} + H_l^T R_l^{-1} H_l \quad (11)$$

The illustration of the RHKF recursion is shown in Fig. 2.

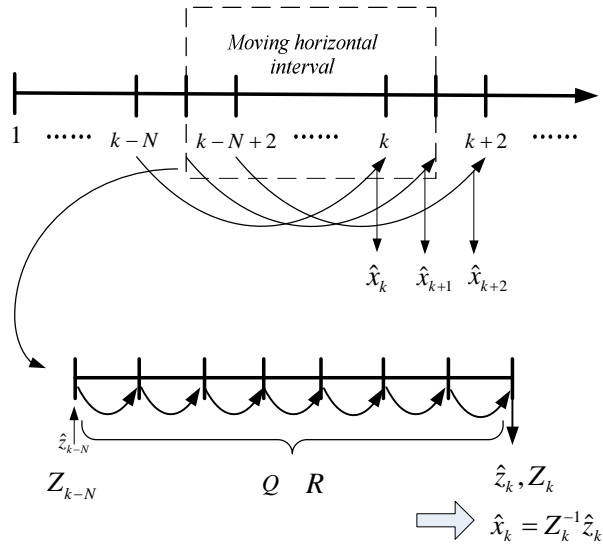

Fig. 2. Illustration of the RHKF Recursion

*C. UFIR in Predictor/Corrector Structure*

The UFIR is an iterative Kalman-Like algorithm ignoring noise and initial conditions. The original Kalman-Like UFIR actually does not have the explicit predictor/corrector structure, however, the predictor/corrector form of the UFIR can be readily obtained as follows

Given a specified horizontal interval $N$ and a certain moving horizontal interval $[k-N,k]$.

Firstly, making use of the measurements and the stacked matrices in interval $[k-N, k-N+n]$ to derive the state estimate at $(k-N+n)$ as $\hat{x}_{k-N+n}$ and $G_{k-N+n}$.

For $l = k-N+n+1, k-N+n+2, \cdots, k$

The prediction step *is*

$$\hat{x}_{l|l-1} = F_l \hat{x}_{l-1} \quad (12)$$

$$G_{l|l-1} = F_l G_{l-1} F_l^T \quad (13)$$

The correction step *is*

$$K_l = G_l H_l^T \quad (14)$$

$$\hat{x}_l = \hat{x}_{l|l-1} + K_l \left( y_l - H_l \hat{x}_{l|l-1} \right) \quad (15)$$

$$G_l = \left( G_{l|l-1}^{-1} + H_l^T H_l \right)^{-1} \quad (16)$$

The illustration of the UFIR recursion is shown in Fig. 3.

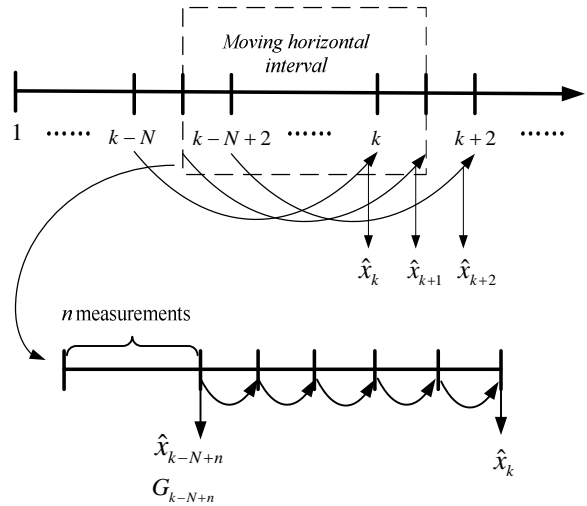

Fig. 3. Illustration of the UFIR Recursion

*D. Discussion*

**Remark 1:** Based on the predictor/corrector form of the iterative RHKF, it can be found that the UFIR can be directly obtained from RHUF by ignoring noise statistics if the horizon initial state is not considered. In the filtering recursion of UFIR, the matrix $G_l$ plays an analogous role as the $P_l$ in RHKF. Through comparison between the filtering equations of RHKF and UFIR, we can also observe how the noise statistics are ignored in UFIR. Since the RHKF and UFIR are both typical representations of FIR filters as shown in Fig. 2 and 3, they possess the same virtues of bounded input/bounded output (BIBO) stability and robustness against temporary model uncertainties, and round-off errors.

**Remark 2:** The ignoring noise statistics in UFIR necessitate the determination of the optimal horizontal interval $N_{opt}$ to minimize the mean square errors (MSE). When $N < N_{opt}$, noise reduction is inefficient and, when $N > N_{opt}$, an increase in $N$ results in an increase in the estimation bias [16]. In contrast, the increase in $N$ will not deteriorate the performance of RHKF and can make the corresponding performance comparable with the KF due to its KF form at each time step in iteration. However, the RHKF requires the knowledge of noise statistics analogous the KF and therefore, the performance suffers when the assumed noise statistics are incorrect. In this respect, the distinctive property of ignoring noise statistics makes UFIR more robust than RHKF under real-world operating conditions.

**Remark 3:** On the estimation horizontal interval, the RHKF is actually a Kalman estimator. Therefore, many fruitful methods that can make the Kalman estimator more robust can therefore be used to improve the performance of RHKF. Some extensions of the KF to nonlinear problems can also be used to design nonlinear RHKF algorithms, precondition to which is that the corresponding nonlinear information filters should be firstly derived.

**Remark 4:** In [17], Zhao and Shmaliy et al. argue that no iterative form was addressed to optimal FIR filtering and they propose an iterative Kalman-Like optimal FIR filter. Since the RHKF is also a type of optimal FIR, the argumentation in [17] seems to be not so appropriate as the iterative form of RHKF has been proposed as far back as 1999. It is shown the developed iterative optimal FIR filter in [8] is as the same form of KF with special initial conditions on the estimation horizon. Since the information filter and the KF are strictly equivalent on the linear state-space, the iterative optimal FIR filters in [8] and [17] should also be equivalent if the initialization of RHKF is determined appropriate. The initializing method of RHKF will be discussed in the next remark.

**Remark 5:** The predictor/corrector form of RHKF (7)-(11) is only used to demonstrate its relationship with UFIR and it can not yet been used directly. This is mainly because that the infinite covariance is employed, which hence can result in a singular problem. Actually, in the original paper of RHKF [8], an alternative information filtering form has be defined irrespective of singularity problems caused by the infinite covariance of the horizon initial state. That form is limited in the assumption that the state transition matrix is invertible, which, however, may be not satisfied in some practical problems. It is known that the information matrix $Z_l > 0$ for all $l \geq n$ with $n$ denoting the state dimension. In this respect, based on the explicit relationship between RHKF and UFIR, the initial state estimation construction procedure using the original $n$ measurements of the horizontal interval in UFIR can be used to initiate the information filter embedded in the RHKF.

**Remark 6:** The relationship between the RHKF and UFIR revealed in this note can benefit the performance improvement of one by drawing lessons from the other. The aforementioned initial condition construction for RHKF is just an example. Moreover, the optimal averaging interval determining method via measurement in a "learning" cycle in UFIR can also afford lessons for RHKF. As is known that the FIR filter with horizontal interval $N$ is about $N$ times slower than KF. In this respect, we would like to determine the interval $N$ that is adequate and not too large. The term "adequate" means that certain most recent measurements should be processed to guarantee the noise denoising effect. Since the increase in $N$ will not deteriorate the performance of RHKF, the determined interval using the method in UFIR can be used as a minimal interval for RHKF.

III. Conclusion

This note systemically compared the RHKF and UFIR —two representative FIR filters and revealed their explicit relationship. This study is expected to facilitate the selection of appropriate filtering algorithms in practice and the development of advanced algorisms by drawing lessons from the virtue of analogous algorithms.